\documentclass [a4paper]{article}
\usepackage{graphicx}
\title{Propagation-based x-ray phase-contrast imaging with broad focus conventional x-ray sources}
\author{M. G. H\"onnicke$^1$, E. M. Kakuno$^2$, J. Manica$^3$\\
and C. Cusatis$^1$}

\begin{document}
\maketitle

\noindent $^1$Departamento de F\'isica, Universidade Federal do
Paran\'a,
Caixa Postal 19091, 81531-990, Curitiba-PR, Brazil\\
$^2$Universidade Federal do Pampa, 96412-420 Bag\'e-RS, Brazil\\
$^3$Departamento de Engenharia Mec\^anica, Universidade Federal
do Paran\'a, Caixa Postal 19011, 81531-980 Curitiba-PR, Brazil

\begin{abstract}

A propagation-based x-ray phase-contrast imaging (PBI) setup
using a conventional x-ray source (LFF Cu target) is presented.
A virtual x-ray source of 40 x 50 $\mu$$m^2$ was created by
using, horizontally, a $6^o$ take-off angle (with the x-ray
tube working in the line focus geometry) and, vertically, a 50
$\mu$m slit . The sample was set 12 $m$ from the source.
Propagation-based x-ray phase-contrast (PB) image and
conventional radiography (CR) of a polypropylene tube were
acquired. Edge enhanced effects and a crack, not detected in
CR, were clearly seen in the PB image. Contrast, visibility of
the object edges and signal to noise ratio of the acquired
images were exploited. The results show that PB images can be
acquired by using normal focus (macro focus) conventional x-ray
sources. This apparatus can be used as an standard
phase-contrast imaging setup to analyze different kind of
samples with large field of view (75 x 75 $mm^2$), discarding
the use of translators for sample and detector.
\end{abstract}

\section{Introduction}
\indent Conventional x-ray radiography is based on the
detection of differences in x-ray attenuation by different
details in an object (sample). Thus, details with similar
attenuation coefficient gives low contrast images. Enhanced
contrast x-ray imaging can be achieved by using attenuation
contrast agents. Alternatively, it can be achieved by
exploiting the real part of the refraction index, which is
responsible for the phase shifts, in addition to the imaginary
part, which is responsible for the absorption. Such
exploitation is done by the well-established x-ray
phase-contrast imaging techniques.\\
\indent Several ways are reported in the literature to acquire
phase-contrast images: a) propagation-based x-ray
phase-contrast imaging (PBI), that uses a
monochromatic~\cite{rf1,rf2} or polychromatic~\cite{rf3}
partially coherent x-ray source; b) x-ray phase-contrast
imaging based on x-ray interferometry~\cite{rf4,rf5} and; c)
analyzer-based x-ray phase-contrast imaging (ABI)~\cite{rf6,
rf7,rf8,rf9,rf10}, that uses diffraction by perfect
single crystals.\\
\indent PBI is well known by its simplicity, once that, it does
not require any sophisticated optics. The unique requirement is
a high quality small source with high brilliance provided by
micro focus x-ray sources~\cite{rf3,rf11,rf12} or, by the high
brilliance and low emittance third generation synchrotron
radiation sources~\cite{rf1,rf2}. This technique can be
developed in the edge detection geometry or in the holographic
geometry~\cite{rf2}, depending on the quality of the source and
the distance from the sample to the detector. Such technique
has been successfully applied severally in biology~\cite{rf13},
medicine~\cite{rf14,rf15} material
sciences~\cite{rf2,rf16} and archeology~\cite{rf17}.\\
\indent In the present work, a PBI setup using normal focus
conventional x-ray source is proposed and realized. The idea is
to get a reasonably transverse coherence length by creating a
virtual micro source and by placing the sample far away from
the source. To characterize such a setup attenuation contrast
images (conventional radiography, CR) and propagation-based
(PB) x-ray phase-contrast images of a polypropylene tube (low
density material) were acquired. The images were also analyzed
in terms of exposure time, contrast and signal to noise ratio.\\
\indent A description of the experiment followed by
quantitative results and conclusions will be presented.

\section{Experiment}
\indent The experimental setup (fig. 1) was mounted using a
normal focus conventional x-ray source (Cu LFF target) set in
the line focus geometry with a take off angle of $6^o$. Such
take-off angle provides a virtual horizontal x-ray source size
of 40 $\mu$$m$. To have an x-ray source with similar vertical
dimension, a 50 $\mu$m slit was set horizontally, just after
the beryllium window of the x-ray tube resulting in a virtual
source size of 40 x 50 $\mu$$m^2$. A vacuum path, 12 $m$ long,
was used along the x-rays path until the sample. Such vacuum
path was made employing PVC tubes with 25 $\mu$$m$ thick kapton
windows and a pressure of 1 mbar. The field of view at the
sample position was 75 x 75 $mm^2$. The measured divergence of
the beam on the sample was $6.10^{-3}$ $rad$ in
the horizontal and vertical scattering planes.\\
\indent The x-ray generator was set at 25 $kV$ in order to have
lower energy photons therefore, large transverse coherence
length, once that, the transverse coherence length ($l_t$) is:

\begin{equation}
\label{eq1} l_t = \frac{\lambda.D}{2.\sigma_x}
\end{equation}

\noindent where $\lambda$ is the wavelength of the incoming
x-ray beam, $D$ is the distance from the source to the sample
and
$\sigma_x$ is the source size.\\
\indent The images were acquired using low-resolution
commercial x-ray films (Kodak INSIGHT dental film). The PB
images were obtained with the x-ray film set $2$ $m$ from the
sample. Another vacuum path was also employed between the
sample and the film. The CR was acquired with the x-ray film in
contact with the sample to avoid phase effects.

\begin{figure}

\centering%
\includegraphics[width=12cm]{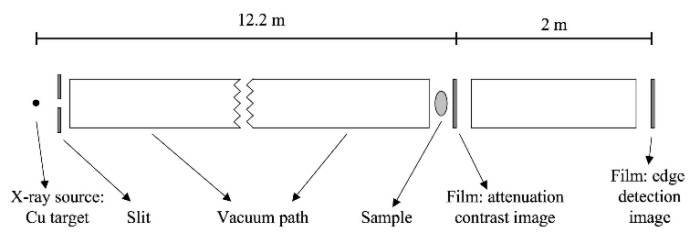}
\caption {Propagation-based x-ray phase-contrast imaging (PBI)
setup. A normal focus conventional x-ray source (Cu LFF target)
was used in the line focus geometry. The take-off angle of
6$^o$ and a horizontal slit of 50 $\mu$$m$ create a
$"$virtual$"$ x-ray micro source of 40 x 50 $\mu$$m^2$. The
sample was set 12 $m$ from the source. Conventional
radiographies (CRs) were acquired with the film in contact with
the sample. Propagation-based (PB) x-ray phase-contrast images
(edge detection images) were acquired with the sample and film
2 $m$ apart.} \label{fig:}

\end{figure}

\section{Results}
\indent To characterize the present PBI setup, a polypropylene
tube, with an external diameter of 6 $mm$ and internal diameter
of 3.8 $mm$ (tube wall 1.1 $mm$ thick) was used as sample. This
tube was fixed on an one-layer paper (about 50 $\mu$$m$ thick).
The CR and PB images of the polypropylene tube are shown in
figs. 2a and 3a, respectively. The cross-sections are shown in
figs. 2b and 3b, respectively. The cross-sections simulations
were done considering an incoming x-rays monochromatic (14.4
$keV$) plane wave beam being attenuated by the sample. A
theoretical pixel size of 25 x 25 $\mu$$m^2$ was considered.
This is far of the real experiment, however it worked fairly
well for the CR, as can be seen in fig. 2b. This means that,
even using a white beam, the energies around 14 $keV$ are the
major contributors for the contrast in CR. This is reasonable,
once that the simulation for a conventional x-ray Cu target at
25 $kV$ shows a maximum of the Bremmstrahlung spectra at 15.5
$keV$~\cite{rf18}.

\begin{figure}

\centering%
\includegraphics[width=12cm]{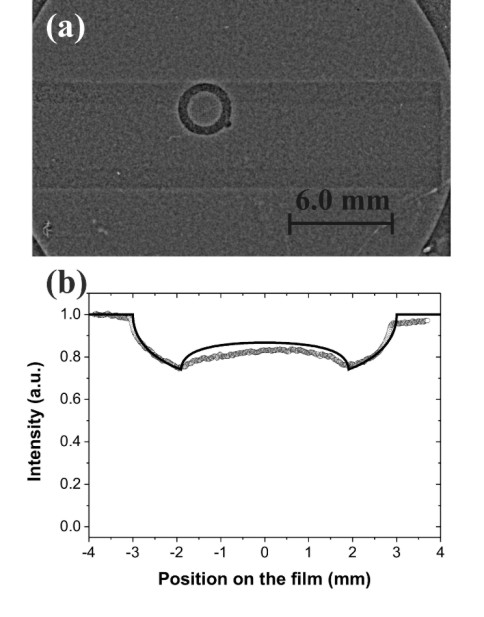}
\caption {(a) Conventional radiography (CR) of a polypropylene
tube. (b) Measured ($o$) and simulated ($-$) cross-section
profiles of the image.} \label{fig:}

\end{figure}

A disagreement between the measured and simulated cross section
profile can be seen for the PB image because the phase effects
and the divergence of the beam were not considered in the
simulations. However, it is worth noticing that phase effects
could be seen in the measured cross-section profiles (indicated
by arrows, in fig. 3b). Moreover, the edges and details of the
tube are better defined in the PB image (fig. 3a) than in the
CR (fig. 2a): for instance, a crack, not visible in the CR, is
clearly seen in the PB image. Also, a background structure can
be noted in the fig. 3a. Such structures can be attributed to
the phase-contrast effects due to the one layer paper behind
the tube. The edge enhanced effect can be attributed to the
rapid variations in the refractive index (boundary effect) that
produces strong phase-contrast, even using a polychromatic
beam, as previously described by Wilkins~\cite{rf3}.\\
\indent For quantitative reasons the images, shown in figs. 2a
and 3a, were compared with each other by measuring the area
contrast ($C$), the signal to noise ratio in the area case
($SNR_{area}$), the visibility of the object edges ($V$) and
the signal-to noise ratio for the edge case ($SNR_{edge}$).
These quantities are defined according to Pagot~\cite{rf19} and
references therein in the following way:

\begin{equation}
\label{eq2} C = \frac{<I_{obj}>-<I_{backg}>}{<I_{backg}>}
\end{equation}

\begin{equation}
\label{eq3} SNR_{area} =
\frac{<I_{obj}>-<I_{backg}>}{\sqrt{\sigma_{obj}^2+\sigma_{backg}^2}}
\end{equation}

\begin{equation}
\label{eq4} V = \frac{I_{max}-I_{min}}{I_{max}+I_{min}}
\end{equation}

\begin{equation}
\label{eq5} SNR_{edge} =
\frac{I_{max}-I_{min}}{\sqrt{2}.\sigma_{backg}}
\end{equation}

\noindent where $<I_{obj}>$ and $<I_{backg}>$ are the mean
intensity values of a given area in the object and in the
background, respectively; $\sigma_{obj}$ and $\sigma_{backg}$
are the standard deviations of the distributions of $I_{obj}$
and $I_{backg}$; and finally, $I_{max}$ and $I_{min}$ are the
maximum and minimum of the mean intensity profile across the
edge.\\
\indent In general, the different PB and CR images were
acquired with different beam intensities, but the exposure time
for the object and for the background were the same.

\begin{figure}

\centering%
\includegraphics[width=12cm]{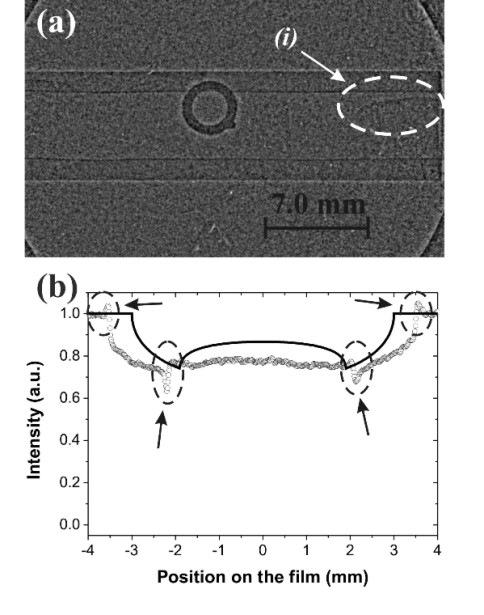}
\caption {(a) Propagation-based (PB) x-ray phase-contrast image
(edge detection image) of the same polypropylene tube shown in
fig. 2. Edge enhancement and a crack (i), not seen in the fig.
2a, are clearly seen here. (b) Measured ($o$) and simulated
($-$) cross-section profile of the image.} \label{fig:}

\end{figure}

\indent The results are shown in tab. 1. The contrast ($C$) and
the visibility of the object edges ($V$) for PB images are
really higher than for CR. However, the good values of V show
that the ABI is highly sensitive to the sample borders (jump of
the phase in these regions). The low values for the signal to
noise ratio are mainly due to the low intensity found in our
measurements. This means, in equations (3) and (5), high values
of  $\sigma_{obj}$ and $\sigma_{backg}$. Such values could be
improved by increasing the exposure times (typically 30
minutes, in our experiment). Also, $SNR_{area}$ and
$SNR_{edge}$ have worse (small) values for the PB images than
for the CRs. A reason for that is because the background was
taken with the paper, which produces phase effects. Therefore,
higher values for $\sigma_{obj}$ and $\sigma_{backg}$ are
found. The present results show that such setup can be used for
phase-contrast x-ray imaging and can be applied for studies of
different kind of samples with large field of view (in the
present work, 75 x 75 $mm^2$). However, this setup can be
improved by using a rotating anode tube (in order to get, at
least, 10 times more intensity), increasing the distance from
the source to the sample and employing a tomography stage with
a high spatial resolution CCD
detector.\\

\begin{center}
\begin{tabular}{|c| ccc r|}
	\hline
Technique &  Contrast($C$)    &   $SNR_{area}$  & Visibility($V$)   & $SNR_{edge}$  \\
	\hline
CR   & -0.19 & -2.66 & 0.31  & 7.52 \\
PBI  & -0.25 & -1.43 & 0.57  & 5.17\\
	\hline
\end{tabular}
\end{center}
Table 1: Contrast ($C$), signal-to-noise ratio in the area case
($SNR_{area}$), visibility of the object edges ($V$) and signal
to noise ratio for the edge case ($SNR_{edge}$) for PB image
and CR. The visibility ($V$) and the $SNR_{edge}$ were obtained
at the top edges of the tube (across the tube wall).

\section{Conclusions}
\indent A propagation-based x-ray phase-contrast imaging setup
(PBI) was mounted using a conventional x-ray source. A virtual
micro focus source was demonstrated by using a $6^o$ take-off
angle and a horizontal slit (50 $\mu$m). The sample was set 12
$m$ from the source. The propagation-based (PB) x-ray
phase-contrast images were acquired with the film and the
sample set 2 $m$ apart. Edge detection effects (that improve
the contrast in the images) and details, such as a crack, not
detected in the attenuation radiography (conventional
radiography, CR) show that a PBI setup with a normal focus
conventional x-ray source, as shown here, can be used as
standard x-ray phase-contrast imaging setup for studying
different kind of samples, with large field of view.
\\
\\
The authors are grateful to PRONEX/CNPq/Fundacao Araucaria and
CNPq for the financial support. M.G. H\"onnicke is grateful to
CNPq/PDJ for the fellowship. The authors also acknowledge
Douglas S.D. da Silva, Hilton C. Guimaraes and Rubens C. da
Silva for the workshop assistance.


\begin{thebibliography}{99} %

\bibitem{rf1}  A. Snigirev, I. Snigireva, V. Kohn, S.
    Kuznetsov, I. Schelokov: Rev. Sci. Instrum. \textbf{66} (1995)
    5486.

\bibitem{rf2}  P. Cloetens, M. Pateyron-Salome, J. Y. Buffiere,
    G. Peix, J. Baruchel, F. Peyrin, M. Schlenker: J. Appl.
    Phys. \textbf{81} (1997) 5878.

\bibitem{rf3}  S. W. Wilkins, T. E. Gureyev, D. Gao, A. Pogany,
    A. W. Stevenson: Nature \textbf{384} (1996) 335.

\bibitem{rf4}  M. Ando and S. Hosoya: Proc. 6th International
    Conference on X-ray optics and Microanalysis ed G. Shinoda
    et al (Tokyo): University of Tokio Press, Tokio (1972) p.
    63.

\bibitem{rf5}   F. Pfeiffer, T. Weitkamp, O. Bunk, C. David:
    Nature Physics \textbf{2} (2006) 258.

\bibitem{rf6}   E. F\"orster, K. Goetz, P. Zaumseil: Krist.
    Tech.
    \textbf{15} (1980) 937.

\bibitem{rf7}   K. M. Podurets, V. A. Somenkov, S. Sh.
    Shil´shtein: Sov. Phys. Tech. Phys. \textbf{34} (6) (1989) 654.

\bibitem{rf8}   V. N. Ingal, E. A. Beliaevskaya: J. Phys. D
    \textbf{28} (1995) 2314.

\bibitem{rf9}   T. J. Davis, D. Gao, T. E, Gureyev, A. W.
    Stevenson, S. W. Wilkins: Nature \textbf{373} (1995) 595.

\bibitem{rf10}  D. Chapman, W. Thomlinson, R. E. Johnston, D.
    Washburn, E. Pisano, N. Gmür, Z. Zhong, R. Menk, F.
    Arfelli, D. Sayers Phys. Med. Biol. \textbf{42} (1997) 2015.

\bibitem{rf11}  J. Jakubek, C. Granja, J. Dammer, R. Hanus, T.
    Holy, S. Pospisil, R. Tykva, J. Uher, Z. Vykydal: Nucl.
    Instrum. Meth. A \textbf{571} (1-2) (2007) 69.

\bibitem{rf12}  T. Tanaka, C. Honda, S. Matsuo, K. Noma, H.
    Ohara, N. Nitta, S. Ota, K. Tsuchiya, Y. Sakashita, A.
    Yamada, M. Yamasaki, A. Furukawa, M. Takahashi, K. Murata:
    Invest. Radiology \textbf{40} (7) (2005) 385.

\bibitem{rf13}  M. W. Westneat, O. Betz, R. W. Blob, K.
    Fezzaa, W. J. Cooper, W. -K. Lee: Science \textbf{299} (2003) 558.

\bibitem{rf14}  R. E. Johnston, D. Washburn, E. Pisano, C.
    Burns, W. C. Thomlinson, L. D. Chapman, F. Arfelli, N. F.
    Gmur, Z. Zhong, D. Sayers: Radiology \textbf{200} (3) (1996) 659.

\bibitem{rf15}  J. Li, Z. Zhong, R. Lidtke, K. E. Kuettner, C.
    Peterfy, E. Aliyeva, C. Muehleman: Journal of Anatomy
    \textbf{202} (5) (2003) 463.

\bibitem{rf16}  J. Baruchel, J. Y. Buffiere, P. Cloetens, M. Di
    Michiel, E. Ferrie, W. Ludwig, E. W. Maire, L. Salvo: Scripta
    Materialia \textbf{55} (1) (2006) 41.

\bibitem{rf17}  Y. Chaimanee, D. Jolly, M. Benammi, P.
    Tafforeau, D. Duzer, I. Moussa, J. J. Jaeger: Nature
    \textbf{422} (2003) 61.

\bibitem{rf18}  R. Jenkins, R. L. Snyder: Introduction to X-ray
    Powder Diffractometry (Chemical Analysis, vol.138) ed J. D.
    Winefordner (John Wiley and Sons) (1996) p. 4.

\bibitem{rf19}  E. Pagot, S. Fiedler, P. Cloetens, A. Bravin,
    P. Coan, K. Fezzaa, J. Baruchel, J. H\"artwig: Phys. Med.
    Biol. \textbf{50} (2005) 709.

\end{thebibliography}
\end{document}